\documentclass[]{aa}
\usepackage{graphicx}
\usepackage{color}%, deluxetable}
\usepackage{natbib}

\begin{document}

%\thesaurus{08(08.1.2, 08.09.2, 08.12.1, 08.13.1, 08.19.6)}
\title{Constraining the age of young stellar clusters via the amplitude of photometric variability }
\author{S.\,Messina\inst{1}}
\offprints{Sergio Messina}
\institute{INAF-Catania Astrophysical Observatory, via S.Sofia, 78 I-95123 Catania, Italy \\
\email{sergio.messina@inaf.it}
}

\date{}
\titlerunning{Age determination from photometric variability}
\authorrunning{S.\,Messina et al.}
\abstract {The determination of stellar age is a crucial task in astrophysics research. At present, the various methods employed in such studies are either model-dependent or based on calibrated empirical relations. The most reliable results are
generally obtained when different methods are applied in a complementary manner. }{We propose a new method for the 
age determination of young stellar associations and open clusters (ages $\la$ 125\,Myr), which may allow for the placement of further constraints on the age
when used in tandem with other methods.} {We explore the amplitude of the photometric variability in bins of color and rotation period of five young associations and clusters spanning an interval of ages from  $\sim$1--3\,Myr to $\sim$625\,Myr (\object{Taurus},  \object{$\rho$ Ophiuchi,} \object{Upper Scorpius,} \object{Pleiades}, and \object{Praesepe}), which all have high-quality time-series photometry from Kepler K2 campaigns.} {In the low-mass regime, we find that stars with similar  color  and rotation periods but different ages exhibit a range of  amplitudes for their photometric variability, with younger stars showing a larger photometric variability than older stars. } {The decline of photometric variability amplitude versus age in stars with similar color   and rotation period may, in principle, be calibrated and adopted as an additional empirical relation for constraining the age of young associations and open clusters, provided that time-series photometry is available for their low-mass members. }
\maketitle
\rm

\section{Introduction}
\indent
Stellar age is a fundamental property in numerous astrophysical contexts. Different methods exist for studies of age determination, such as astero-seismology, or the comparison of measurable stellar parameters with stellar evolutionary models (for example, isochrone and lithium-depletion boundary methods), or with calibrated empirical relations (e.g., gyro-chronology, 
specific element abundance ratios, activity proxies). However, the latter requires firm calibration and any other method is better suited for  limited regions of the parameter space, making age determination a particularly challenging task (see \citealt{Soderblom10} for a review). 

The most reliable results are generally obtained when different methods are applied in a complementary manner (see, e.g., \citealt{Desidera15}). From this perspective, we intend to explore a new empirical relation for the age determination, which is based on the age dependence of the level of photometric variability exhibited by low-mass young stars and originating from magnetic activity. 

To avoid confusion, it is worth noting that in the present study, we do not explore the variation of the level of photometric variability versus time arising from the decrease in magnetic activity that accompanies the slowdown of the rotation as stars age (the well-known activity-rotation connection;  see, e.g., \cite{Rosner80}, \cite{Catalano91},  \cite{Messina03}). 
Rather,
we show evidence that young low-mass single stars with similar color and similar rotation period can exhibit significantly different levels of photometric variability at different ages.\\
\indent
The photometric variability observed in post T-Tauri single low-mass stars is believed to be entirely produced by the magnetic activity on the stellar photosphere (\citealt{Schrijver00}). Intense magnetic fields at photospheric levels, which manifest as dark and bright spots, induce variations of the stellar flux over a range of time scales and produce photometric variability (see, e.g., \citealt{Messina04}).
The level of photometric variability depends on the level of magnetic activity which, in turn, depends on the efficiency of the underlying hydromagnetic processes that take place in the stellar interior.\\
\indent
In solar-mass stars that are partially convective,   magnetic activity is explained as originating from an $\alpha\Omega$ dynamo whose efficiency is driven by rotation (differential rotation) and convection turnover time, which is related to the depth of the convection zone (see, e.g., \citealt{Parker79}; \citealt{Rosner80}; \citealt{Schuessler83};  \citealt{Weiss94}). The faster the rotation and the deeper the convection zone, the larger the dynamo efficiency, the total amount of surface magnetic fields, and the observed variability. A number of authors have quantified the dependence of various activity indices on rotation and mass (e.g., \citealt{Messina03}; \citealt{Mittag18}; \citealt{Brun17}).\\
\indent
In lower-mass stars that are  almost fully convective, magnetic activity is explained as originating from an $\alpha^2$ dynamo whose efficiency is driven by turbulence (see, e.g. \citealt{Brandenburg05}).  Interestingly, similar rotation-activity relationships exist  in both mass regimes (\citealt{Wright18}).\\
\indent
The decline of magnetic activity with age, which was first quantified by Skumanich (1972), has been attributed to the slowing down of the rotation with age (owing to the effects of braking by magnetized stellar winds). The slowing down of stellar rotation  determines a decrease in the dynamo efficiency and a consequent decrement of magnetic activity and of photometric variability.

In the present study,  we show evidence that young stars sharing similar  color  and rotation, but different ages indeed show different levels of photometric variability that declines with age. For example, a 3 Myr 0.8-solar mass star with a 5 d rotation period has an amplitude of photometric variability significantly larger than a star of 8 Myr and a solar mass of 0.8 with the same  5 d rotation period, and that trend holds down to an age of  about 125\,Myr for the same mass and rotation period. We investigate the origin of this behavior and whether it arises from a change in the stellar internal structure or in the topology  of surface magnetic fields. 

The dependence of the amplitude of photometric variability on age, at fixed  color  and rotation period, can be exploited to infer a new calibrated empirical relation as a complementary method to constrain the age of young  stellar open clusters and associations, provided that accurate measurements of variability of their low-mass members become available from high-precision photometric time series. 

In Sect.\,2, we describe our working data sample. In Sect.\,3 we present our analysis. In Sect. 4, we present our discussion and give our conclusions in Sect.\,5.

\begin{figure}
\begin{minipage}{10cm}
%\centering
\includegraphics[scale = 0.31, trim = 10 10 0 40, clip, angle=90]{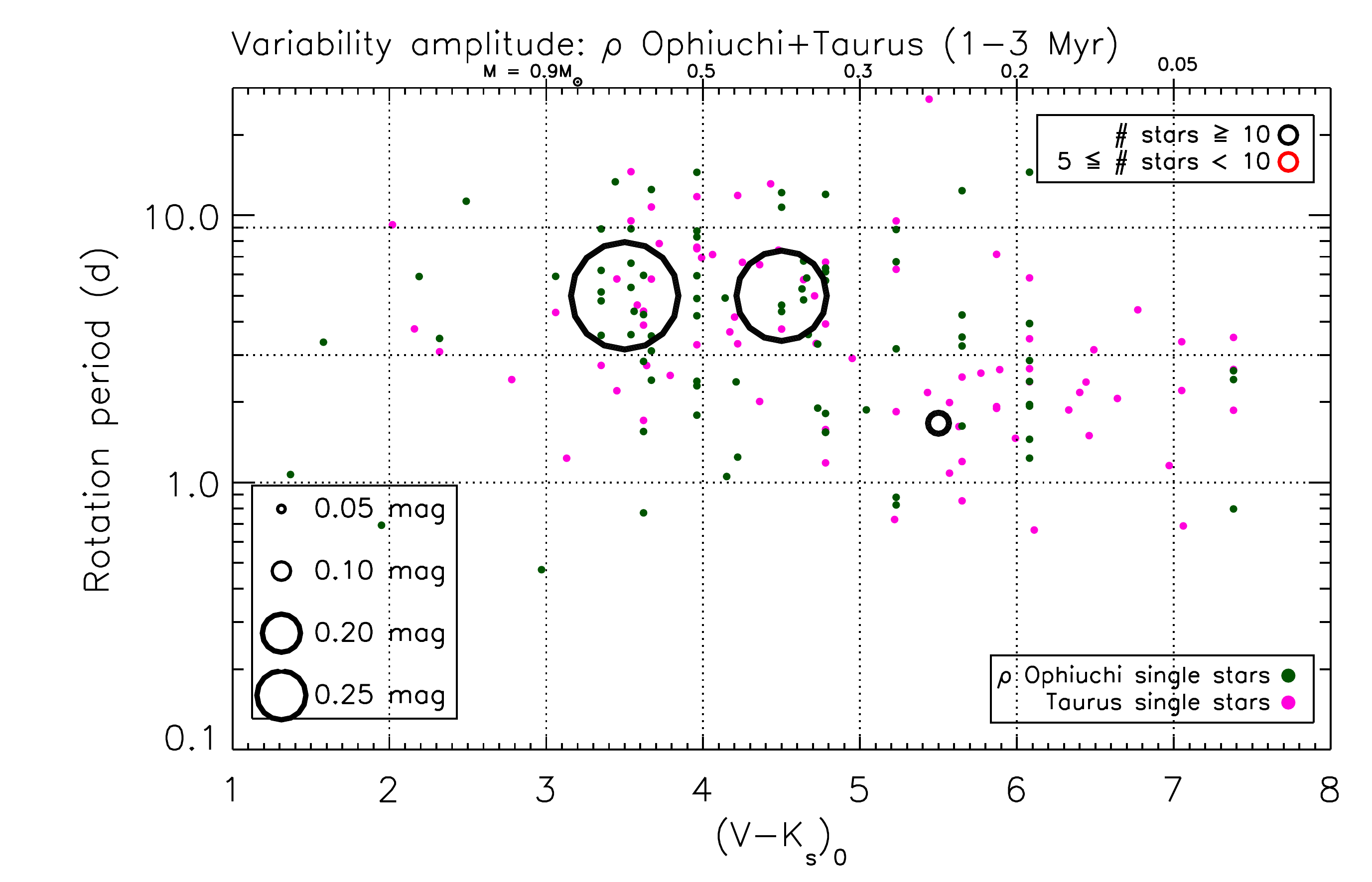}\\
\includegraphics[scale = 0.31, trim = 10 10 0 40, clip, angle=90]{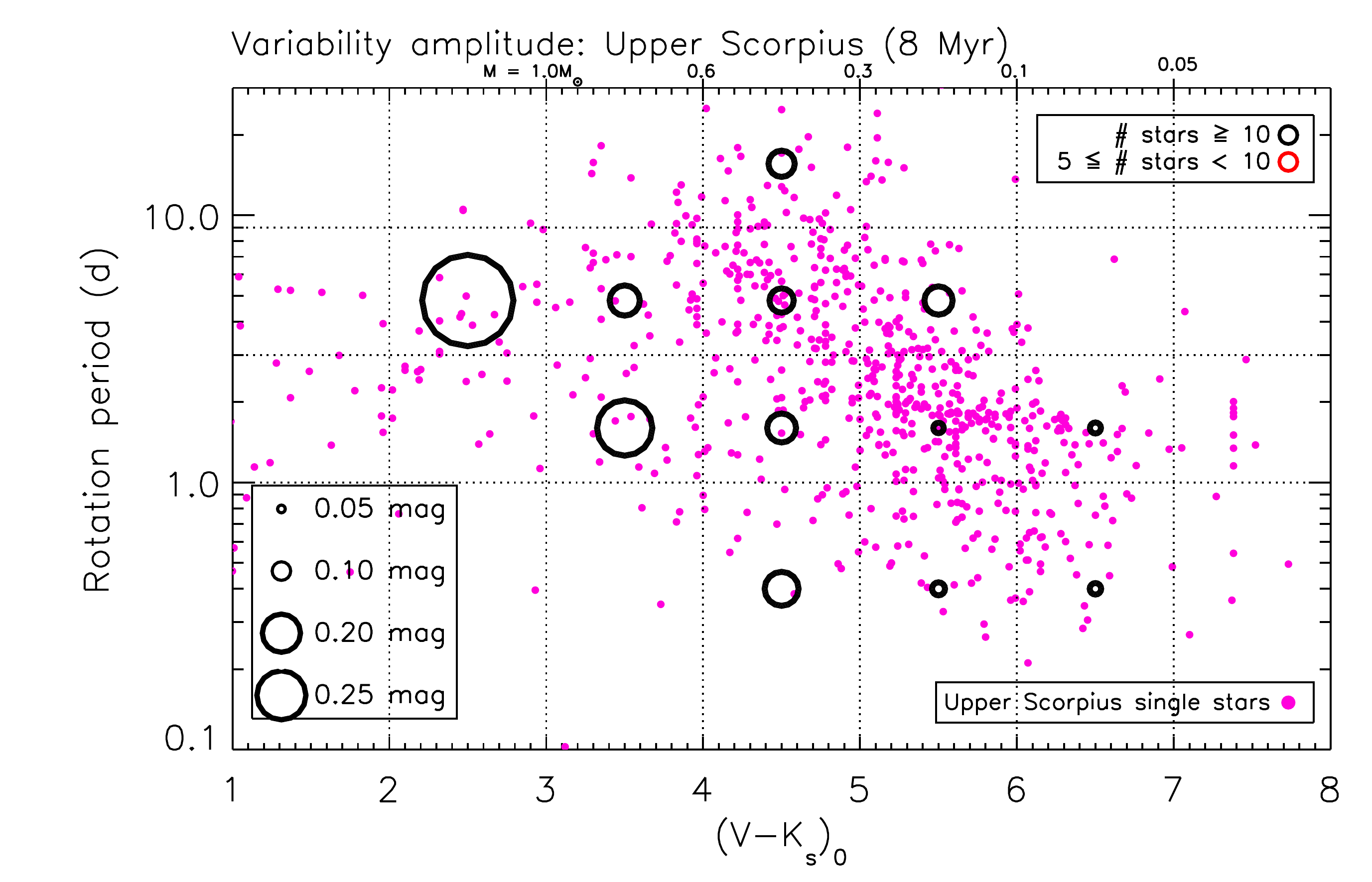}\\
\includegraphics[scale = 0.31, trim = 10 10 0 40, clip, angle=90]{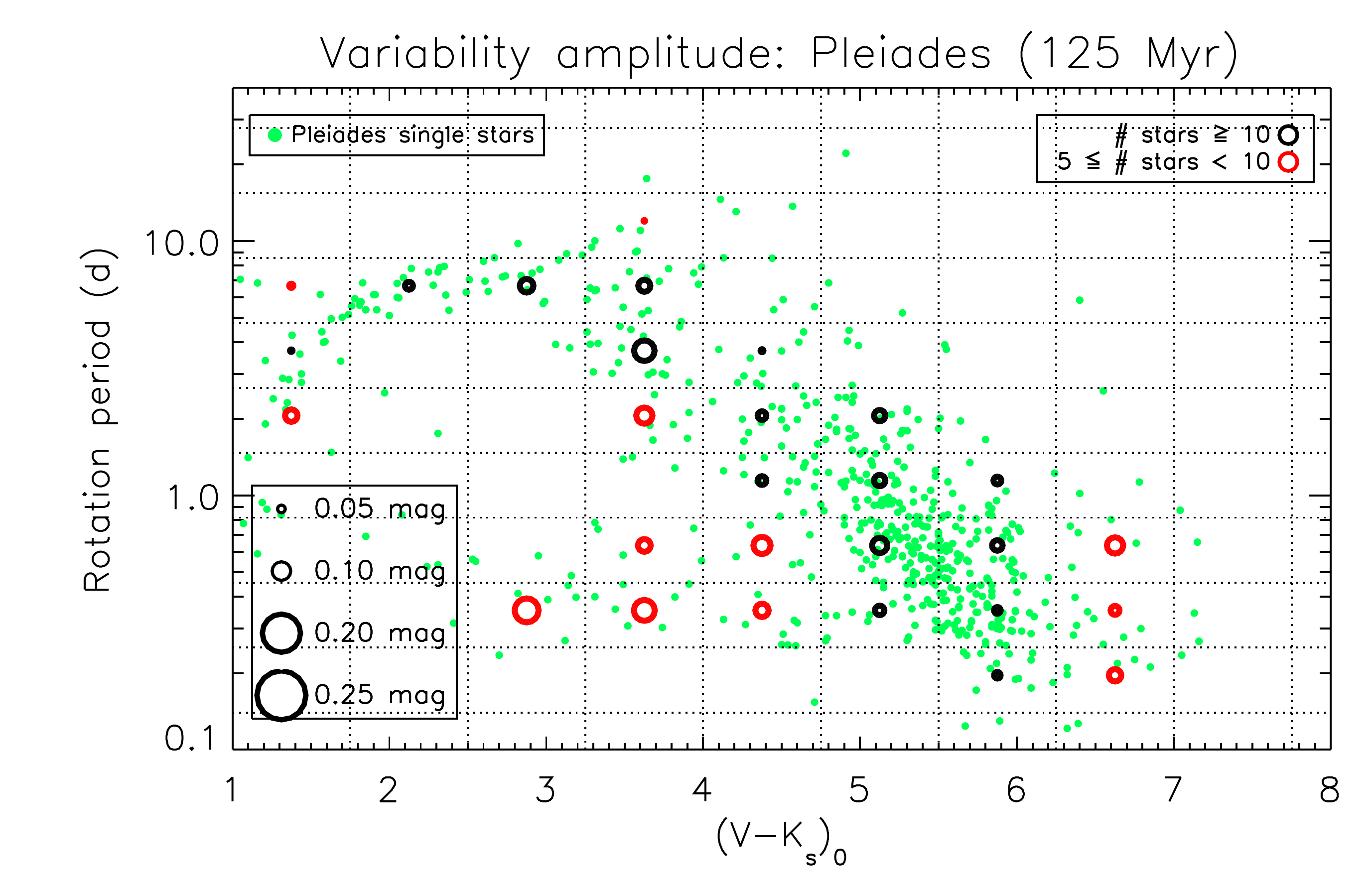}\\
\includegraphics[scale = 0.31,trim = 10 10 10 40, clip, angle=90]{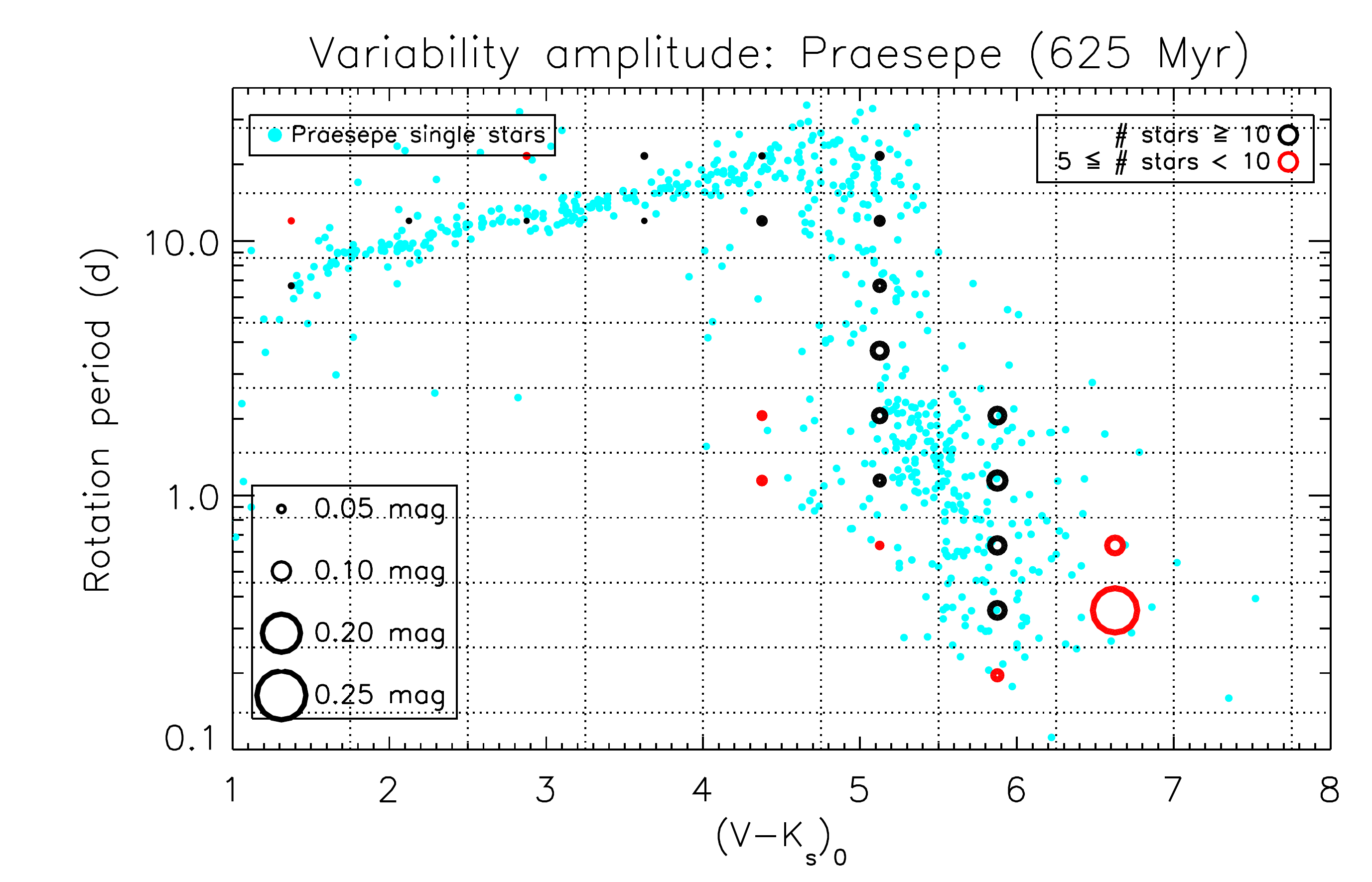}\\
\end{minipage}
%\vspace{1cm}
\caption{\label{color-period} Distribution of stellar rotation periods versus dereddened color for candidate members of the  Taurus (\citealt{Rebull20}),  $\rho$ Ophiuchi and Upper Scorpius associations (\citealt{Rebull18}), Pleiades  (\citealt{Rebull16}), and Praesepe (\citealt{Rebull17}). Open circles have size proportional to the amplitude of photometric variability (@80th percentile).}
\end{figure}

\section{Data}

For the purposes of this study, we selected five cases among the associations and open clusters of known ages: \object{Taurus} ($\le$3\,Myr; \citealt{Kraus09}), \object{$\rho$ Ophiuchi} ($\sim$2--5 Myr; \citealt{Wilking08}), \object{Upper Scorpius} ($\sim$8\,Myr; \citealt{Feiden16}, \citealt{Herczeg15}), \object{Pleiades} ($\sim$125\,Myr; \citealt{Stauffer98}), and \object{Praesepe} ($\sim$625\,Myr; \citealt{Brandt15}).
In the analysis, we used the stellar rotation periods measured from the ultra-high precision photometric time series collected by Kepler K2  during campaign C13 (for Taurus; \citealt{Rebull20}), C2 (for $\rho$ Ophiuchi and Upper Scorpius; \citealt{Rebull18}), C4 (for Pleiades; \citealt{Rebull16}),    and C5 (for Praesepe; \citealt{Rebull17}).
%In the analysis of the Praesepe cluster, we used the stellar rotation periods collected by \citet{Douglas17}, who complemented their own period measurements from Kepler K2 light curves (campaign C5) with rotation periods from \cite{Kovacs14}, \cite{Delorme11}, \cite{Scholz07}, \cite{Scholz11}, and  \citet{Agueros11}.
We gathered the dereddened (V$-$K$_s$)$_0$ colors from the aforementioned papers, respectively, by Rebull et al.. \rm
%of the \bf Taurus, \rm $\rho$ Ophiuchi and Upper Scorpius associations and Pleiades \bf  from \cite{Rebull20}, \rm \cite{Rebull18} and  \cite{Rebull16}, whereas for Praesepe (whose reddening is negligible)(V$-$K$_s$) colors were gathered from \citet{Rebull17}.
%Rotation periods of the members of  the Praesepe cluster were taken from the compilation of \citet{Douglas17}. The \citet{Douglas17} catalogue consists of 685 rotation periods of which 
%471 were measured from Kepler K2 light curves collected during the C5 campaign; 107 period measurements were taken from \cite{Kovacs14}; 36 measurements from \cite{Delorme11}; 35 from \cite{Scholz07}\citeyear{Scholz11}; and 36 from \citet{Agueros11}.
%In our analysis, we considered only the 592 rotation periods of the single stars, excluding confirmed and candidate binaries, as classified by \citet{Douglas17}.\\
%The (V$-$K$_s$) colors of the Praesepe members were taken from \cite{Rebull17}.\\

%In our analysis, we considered only the 601 rotation periods of the single stars, excluding binaries. Dereddened (V$-$K$_s$)$_0$ colors were taken from \citet{Rebull16}.
From the mentioned period databases, we selected only single stars, excluding known and suspected binaries, as well all stars that were found to exhibit multi-periodicity. In fact, the latter are likely unresolved photometric close binaries (see, e.g. \citealt{Stauffer18}, {Messina 2018}).
We focused our analyses solely on single stars because the rotation period evolution of close binaries is expected to be different from that of single stars, owing to the gravitational tidal effects between the system's components. Moreover, while the photometric variability of single stars under analysis arises from magnetic activity,  in binaries, there is the spurious contribution to variability by eclipses and by the ellipsoidal shape of the binary's components. 

 Since the amplitudes of the rotational variability were not provided for all associations and clusters in the mentioned source papers, we computed this quantity in a homogeneous way for all targets in the study. 
For each star in our study, we retrieved the archived Kepler light curves and measured the photometric variability amplitude by transforming the Pre-search Data Conditioning Simple Aperture Photometry (PDCSAP)  fluxes  into magnitudes. We removed long-term linear trends and outliers at 5-$\sigma$ levels. Finally, we measured the amplitude of the 5th to 95th percentile, which we adopted as a measure of the photometric variability amplitude. 
The same procedure was carried out for all the light curves.

\section{Analysis}

In Fig.\,\ref{color-period}, we plot the distribution of the rotation periods, P, versus the dereddened (V$-$K$_s$)$_0$ colors for the low-mass single members of Taurus and $\rho$ Ophiuchi (whose ages are comparable within their uncertainties), Upper Scorpius, Pleiades, and Praesepe. To explore the dependence of the photometric variability amplitude on age
at constant  color  and rotation, we used a grid of values (dotted lines). The grid binning should guarantee a sufficient number of stars within each grid element with which to draw meaningful statistics and a sufficiently fine sampling that considers all stars within each bin of a similar color  and comparable rotation period.
In the color and rotation period range of our target stars,  a grid of  7$\times$4 elements  was found to be the best compromise.  Of a total of  28  grid elements, we considered  only elements containing 15 stars  at least for the subsequent analysis. Since the rotation period distribution changes with the age, not all grid areas are uniformly populated by members of all five clusters and associations. 
 For instance, using a binning similar for the color but finer for the rotation period (totaling  8$\times$9 elements), we found qualitatively similar results, although with a slightly poorer statistics. \\
It is important to recall that the amplitude of the photometric variability depends on several geometrical and physical parameters. Some parameters are constant in time, such as the inclination of the star's rotation axis; or they can be assumed to be constant, such as the brightness contrast between spot and photosphere. Others are variable, such as the total area covered by spots and their surface distribution. The same level of activity generally generates a range of amplitudes of the rotational variability, depending on the values of these parameters (e.g., from $\sim$0 mag for stars seen pole-on  up to a few tenths of magnitude for stars seen  equator-on). Therefore, the photometric variability when used as activity diagnostics may underestimate the true activity level hosted by a star.  \\
%However, the longer the timeseries the higher the probability that the evolution of active regions allows to reveal a value of activity closer to the maximum for that star.\\
\begin{figure}
\begin{minipage}{10cm}
%\centering
    \includegraphics[scale = .55, trim = 0 110 10 50, clip, angle=0]{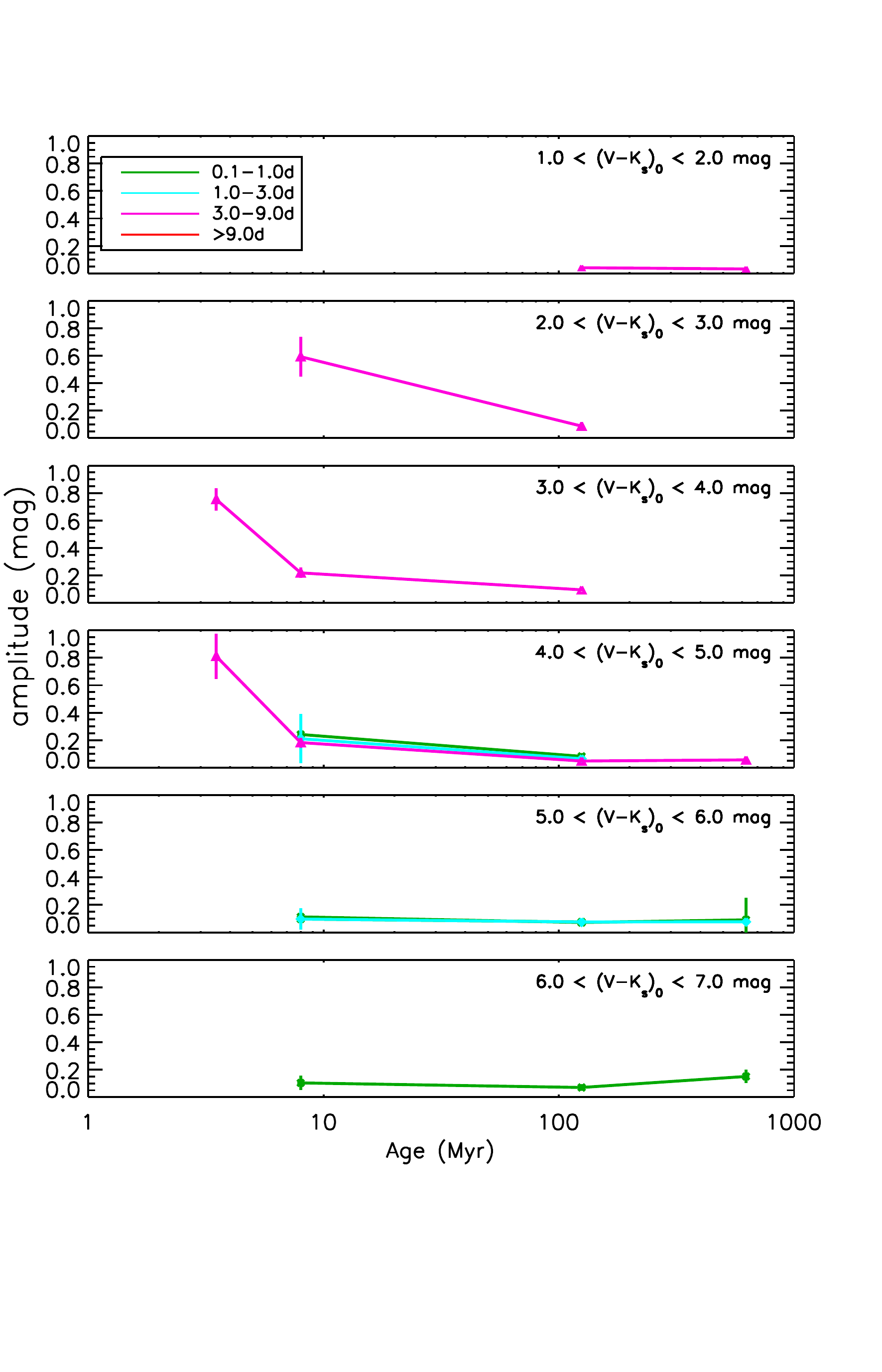}
    \end{minipage}
\caption{\label{age} Amplitude of photometric variability versus age. Stars are divided in six color bins and different colors are used to distinguish among different rotation period bins. Solid lines are used to simply connect  data.}
\end{figure}
On the basis of this consideration, instead of using the average photometric variability amplitude within  each grid element containing  15 stars, at least,  and in order to minimize the effect of possible outliers (i.e. of stars with an activity level beyond normal values), we opted to use the 80th percentile as we find it to be more representative of the true activity level for the given rotation and  color. 
For such grid elements with  15 stars,  at least, in Fig.\,\ref{color-period} we over-plot black open circles whose size is proportional to the  amplitude (@80$^{th}$ percentile) of photometric variability for the Taurus and $\rho$ Ophiuchi (top panel), Upper Scorpius (second panel from top), Pleiades (third panel), and Praesepe members (bottom panel). Numerical values of amplitudes and corresponding uncertainties are listed in Table\,\ref{tab:table}. \\
%Only for illustrative purposes but not for subsequent analysis, we also plot  the amplitudes in those grid areas with a smaller number of stars from 5 to 9 using red open circles.\\
\indent
A visual inspection is enough to make it clear that  the  amplitude of photometric variability  decreases from the age of our youngest associations (Taurus and $\rho$ Ophiuchi) untill the age of our oldest cluster (Praesepe)   at a fixed color  and rotation period in most of the sampled grid areas.\\
\indent
To quantify the age decline of the amplitude of photometric variability, we identified those grid elements with measured amplitude and which are common to all clusters and associations as well as to couples of them. As expected, due to the different period distributions at different ages, we had to limit our analysis to a fraction of the total gridded area.  For instance, in those coincident grid elements, we counted at least 20 stars, which allows for a more statistically significant amplitude measurement. \rm\\
\indent
In Fig.\,\ref{age}, we plot the amplitude of photometric variability versus the four sampled ages  (1--3\,Myr, 8\,Myr, 125\,Myr, and 625\,Myr). Amplitudes corresponding to the same grid area are plotted with the same symbol and connected by solid lines. Different colors are used to distinguish different rotation period bins.

\section{Discussion}
To correctly interpret the derived pattern of variability versus age, it has to be  taken into account that among the single members of the  Taurus and \rm $\rho$ Ophiuchi associations selected for our analysis, about 50\% of them are host to discs (\citealt{Rebull18}, \citeyear{Rebull20}). A fraction of these discs are likely to be accreting onto the star, giving rise to an additional contribution to the observed photometric variability. The accretion process, via the generation of a  hot spot on the star's surface, can produce photometric variability comparable to -- or larger than -- that which originates from magnetic activity (see, e.g., \citealt{Messina16}, \citeyear{Messina17}). On the other hand, discs have been detected in less that 5\% of single members of the Upper Scorpius association that were selected for our analysis (\citealt{Rebull18}). 
 Among members of $\rho$ Oph and Taurus, the number of stars with no evidence for discs is not enough to gather a significant statistics (i.e., less than 15 per bin). Therefore, the amplitudes reported in  Fig.\,\ref{age} at the age of  1--3\,Myr refer only to stars with evidence for discs and it is very likely these arise from accretion phenomena. \\
\indent
Another aspect that needs to be considered is that low-mass stars change their surface temperature and photometric color during their evolution. This is the reason why the color is generally not an accurate mass proxy when stellar properties at very different ages are compared.  However, in our investigation  stars have mostly vertical evolutionary track in the Hertzsprung-Russel diagram, apart from the more massive  members of Upper Scorpius ((V$-$K$_s$)$_0$ $<$ 3\,mag), and  the color variation of our stars versus age is comparable to the grid color resolution. 
These circumstance makes the color a suitable mass proxy for the aim of our study.\\
\indent
In the following, we summarize the observed behavior of amplitude versus age in different color bins.
%\indent
\begin{itemize}
\item \it 1.0 $\le$ (V$-$K$_s$)$_0$ $\le$ 2.0\,mag:  \rm photometric amplitudes are available for stars with  3 $<$ P $<$ 9\,d and ages between Pleiades and Praesepe. No variation of amplitude versus age is detected.\rm
%\indent
\item \it 2.0 $\le$ (V$-$K$_s$)$_0$ $\le$ 3.0\,mag: \rm photometric amplitudes are available for stars with  3 $<$ P $<$ 9\,d and ages between Upper Scorpius and Pleiades. A significant amplitude decrease by a factor $\sim$6 is measured from 8 to 125 Myr.   As anticipated,  Upper Scorpius  are more massive than Pleiades members in this color bin.  
%in this color range we have photometric amplitudes for stars with  P $\ge$ 2\,d. We note a significant decrease of the photometric amplitude from the age of 8\,Myr to the age of 125\,Myr by a factor $\sim$2 (3.7\,d rotation bin) to $\sim$5 (6.5\,d rotation bin). The decrease of amplitude from the age of 125\,Myr to 625\,Myr (sampled only for stars with P $>$ 9\,d) is barely detected.\\
%\indent
\item \it 3.0 $\le$ (V$-$K$_s$)$_0$ $\le$ 4.0\,mag: \rm
 photometric amplitudes are available for stars with  3 $<$ P $<$ 9\,d and ages at Taurus+$\rho$ Ophiuchi, Upper Scorpius, and Pleiades. A significant amplitude decrease by a factor $\sim$4 is measured from 1--3 to 8 Myr, and by a factor 2 from 8 to 125 Myr.
%\indent
\item \it 4.0 $\le$ (V$-$K$_s$)$_0$ $\le$ 5.0\,mag: \rm
we note a significant decrease in the photometric amplitude from the age of 1-3\,Myr to the age of 8\,Myr by a factor $\sim$4  for stars with  3 $<$ P $<$ 9\,d. However, as noted before the variability at 1-3\,Myr is likely dominated by accretion phenomena.
Another decrease in the photometric amplitude is noted from the age of 8\,Myr to the age of 125\,Myr by a factor $\sim$2  similarly for about all rotation period bins, whereas no variation is measured at older age.

\item \it 5.0 $\le$ (V$-$K$_s$)$_0$ $\le$ 6.0\,mag: \rm
 photometric amplitudes are available for stars with  0.1 $<$ P $<$ 3\,d  and the amplitude is about constant from 8 to 650 Myr. 
%\rm in this color range we have photometric amplitudes for stars with  P $<$ 0.6\,d. We note no variation  of the photometric amplitude from the age of 8\,Myr to the age of 125\,Myr  for stars with P $\sim$0.6\,d, and marginal evidence of an increase from the age of 125\,Myr to the age of 625\,Myr.\\
%\indent
\item \it 6.0 $\le$ (V$-$K$_s$)$_0$ $\le$ 7.0\,mag: \rm
 photometric amplitudes are available for stars with  0.1 $<$ P $<$ 1\,d and at ages of Upper Scorpius, Pleiades, and Praesepe. The amplitude is about constant, although there is some hint for an increase from the 125 to 625 Myr.\\
\rm
\end{itemize}
%We note that the amplitude variations derived from grid areas with poor statistics (five stars at least) are in  qualitative agreement with the amplitude variations derived from grid areas with better statistics (more than 10 stars).\\
\begin{figure}
\begin{minipage}{10cm}
%\centering
\includegraphics[scale = .35, trim = 30 0 30 0, clip, angle=90]{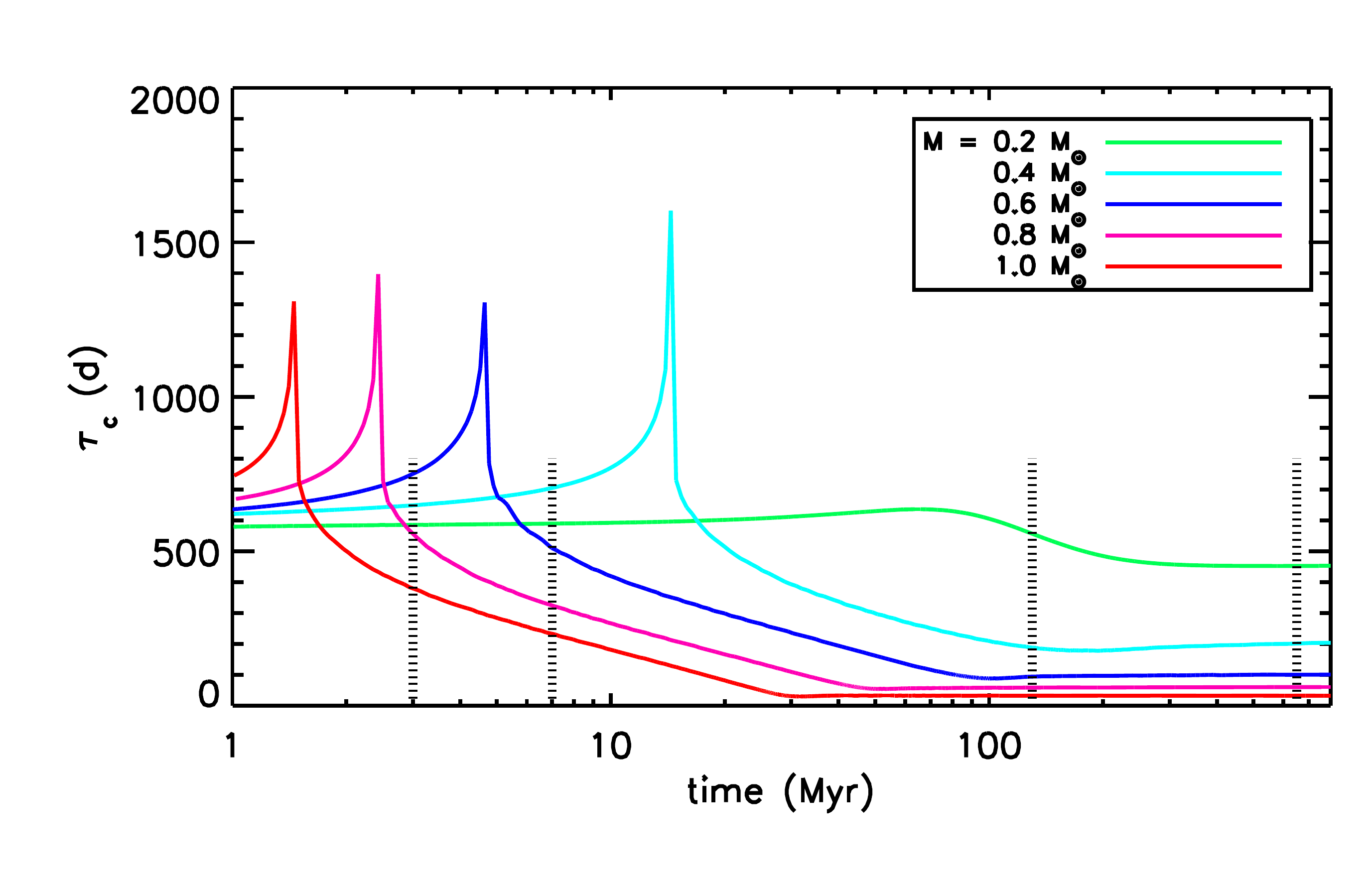}
\end{minipage}
\caption{\label{tau} Convective turnover time from YAPSI evolutionary models versus time in the  0.2--1.0\,M$_{\odot}$  mass range. Vertical dashed lines indicate the age of the associations and opens clusters under analysis.}
\end{figure}
\indent
The general trend that we observe is common to different color bins, that is, a steep decrease in amplitude in the 1--8 Myr age range, followed by a less steep decrease till the oldest sampled age of 625\,Myr. Differently, in the lowest mass regime, (V$-$K$_s$)$_0$ $\ge$ 5.0, the amplitude decrease is barely detected, when, rather, the opposite behaviour seems to be  observed among the reddest and oldest stars in our sample (the amplitude increases by a factor of 2 from the Pleiades to the Praesepe age). \\     
Regarding this point, it is worthwhile recalling that the amplitude of photometric variability due to starspots shows evidence of saturation among the fastest rotators in the sense that the amplitude of the photometric rotational modulation stops increasing at the shortest rotation periods -- even if it does not even show hints of decreasing. This saturation is observed for Rossby numbers R$_0$ $<$ 0.02 or P $<$ 0.35\,d for G--K-type stars (see, \citealt{Messina01}, \citealt{O'dell95}). In our sample, only low-mass (5.0 $<$ V$-$K$_s$ $<$ 7.0\,mag) members of the Pleiades and Praesepe fall in this saturation regime. Therefore, the increase in amplitude from the age of 125 to the age of 625\,Myr, which is detected in our analysis, may even  be underestimated to some extent owing to the starspot saturation and consequent decrease in photometric variability amplitude.

\begin{figure}
\begin{minipage}{10cm}
%\centering
\includegraphics[scale = .35, trim = 0 10 0 60, clip, angle=90]{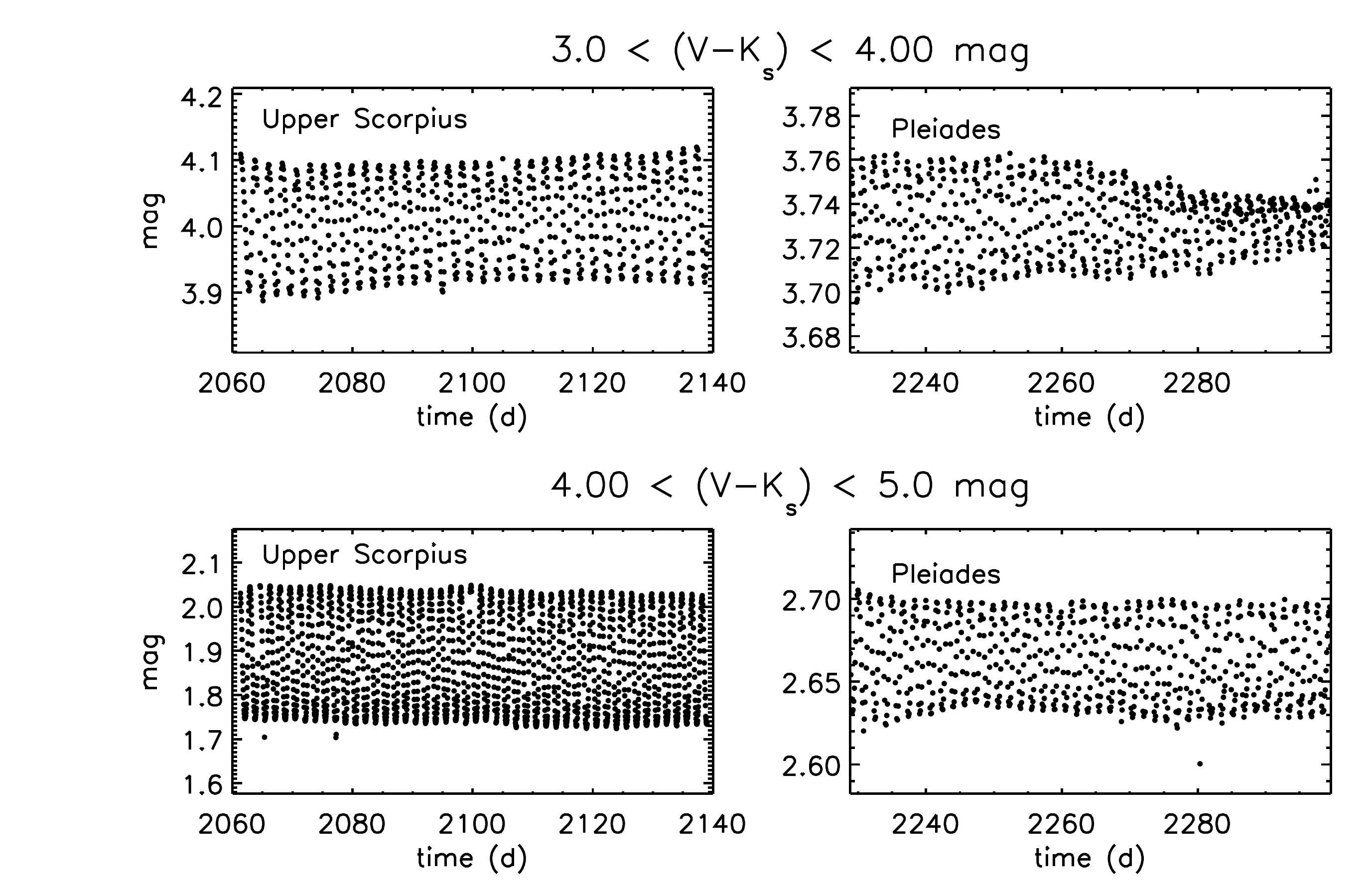}
\end{minipage}
\caption{\label{argd} \it Upper panels: \rm  Examples of time variation of the light curve peak-to-peak amplitude, from stable pattern at the age of Upper Scorpius to much less stable pattern at the Pleiades age in the 3.0 $\le$ (V$-$K$_s$)$_0$ $\le$ 4.00\,mag color range.  \it Bottom panels: \rm  Light curve stability remains mostly unchanged at the Upper Scorpius and Pleiades ages in the 4.00 $\le$ (V$-$K$_s$)$_0$ $\le$ 5.0\,mag color range. }
\end{figure}

\rm
%The color range that could be better explored is 3.25 $\le$ (V$-$K$_s$)$_0$ $\le$ 4.75\,mag, approximately corresponding to the mass interval from $\sim$0.4 to $\sim$0.7\,M$_{\odot}$. In this range, the decline of the amplitude of photometric variability from the age of 8\,Myr to 125\,Myr is significant (also detected from the age of 3\,Myr, but with less significance).  \\
\indent
The evolution over time of the internal stellar structure may be called upon to possibly explain this decreasing trend. A property of the internal structure, which is known to change over time and also to be positively correlated to the dynamo efficiency, is the depth of the external convective envelope, which, in turn, sets the value of the turnover timescale  ($\tau_c$).
In Fig.\,\ref{tau}, we show that $\tau_c$ changes significantly and in a complex way at young ages,  with clear mass dependence. Indeed, despite the temporary rises in the 3--10 Myr range, $\tau_c$ decreases from the age of 1--3\,Myr (Taurus and $\rho$ Ophiuchi) to the age of 8\,Myr (Upper Scorpius) and even further to the age of 125\,Myr (Pleiades) in the mass interval under consideration. In the present study, we retrieved 
$\tau_c$ from the YaPSI (Yale-Potsdam Stellar Isochrones; \citealt{Spada17}) collection of stellar evolutionary models, selecting the solar composition. The decrease in $\tau_c$, which implies a decrease in the dynamo efficiency, seems to reasonably explain the observed decrease in the photometric variability amplitude.  For instance, we notice that the almost constancy versus age of the variability amplitude at the reddest color bin is reflected by a comparable constancy versus age of $\tau_c$ (green line in Fig.\,\ref{tau}). \\ 
\indent
The evolution over time with respect to the topology of the surface magnetic fields could be also called upon to explain the time decrease in the activity level. In this hypothesis,  older stars may tend to have active regions on their surface  distributed more uniformly than do younger stars of similar mass and rotation period, giving rise to lower levels of variability. This also indicates that not only the distribution but also the timescales of the active region growth and decay (ARGD) may be time-dependent.  Stable and highly asymmetric active regions at younger ages may tend to become shorter-lived and more  homogeneously distributed at older ages, making the flux rotational modulation take on a progressively smaller amplitude.\\
\indent
In order to explore this hypothesis on qualitative grounds, we visually inspected all the light curves and noted that in the 3.0 $\le$ (V$-$K$_s$)$_0$ $\le$ 4.0\,mag color range,  Upper Scorpius members have stable light curves, that is, the peak-to-peak amplitude of the flux rotational modulation remains mostly unchanged during the 80 days of the K2 campaign, indicating that the ARGD occurs on relatively long time scales. Differently, the Pleiades members show less stable light curves, whose peak-to-peak amplitude varies from one rotation cycle to the next one. \\
On quantitative grounds,  the standard deviation $\sigma$ of the peak-to-peak amplitude has been computed for each star and then averaged over all stars within the same  area element of the period-color plane (see Fig.\,\ref{color-period}). It has been established that Pleiades members have an average $\sigma$ that is two times larger than USco members, independently from the rotation period, in the 3.0 $\le$ (V$-$K$_s$)$_0$ $\le$ 4.0\,mag color range. In the top panels of Fig.\,\ref{argd}, we provide an example of such different behaviours. Whereas USco members spend most time exhibiting their highest variability amplitude, Pleiades members spend a much shorter time at that level. That increases the probability of observing a Pleiades member at a low-amplitude state that would measure smaller light-curve amplitudes.  \\
\indent
In contrast, in the 4.0 $\le$ (V$-$K$_s$)$_0$ $\le$ 5.0\,mag color range (bottom panels of Fig.\,\ref{argd}), both USco and Pleiades stars show similar average $\sigma$, independently from the rotation period range. Nonetheless, the variability level of USco stars is about two times the level of the Pleiades (see Fig.\,\ref{age}). Therefore, in addition to $\tau_c$,  the evolution over time of the ARGD timescale can be responsible of the observed decreasing trend of the variability amplitude, at least the 3.0 $\le$ (V$-$K$_s$)$_0$ $\le$ 4.00\,mag color range.
% Requires the booktabs if the memoir class is not being used
\begin{table*}[htbp]
   \centering
   \caption{Amplitudes (@80 percentile) of the photometric rotational modulation in bins of V$-$K$_s$ color and rotation period.} % requires the topcapt package
   \begin{tabular}{@{} l|lcccccc @{}} % Column formatting, @{} suppresses leading/trailing space
        \hline
          & V$-$K$_s$\,(mag)    & 1--2 &2--3 & 3--4 & 4--5 & 5--6 & 6--7   \\
          \hline
        \,\,\,P\,(d) &   &&  &  &  &  \\
        \hline
                & \multicolumn{7}{c}{\bf Taurus + $\rho$ Oph}\\
      0.1--1.0   &      &  --- & --- & --- & --- & --- & --- \\
      1.0--3.0   &    &  --- & --- & --- & --- & 0.16$\pm$0.02 & ---  \\
      3.0--9.0& &  --- & --- & 0.75$\pm$0.10$^{\star}$ & 0.80$\pm$0.07$^{\star}$ & --- & --- \\
      9.0--30     &    &  --- & --- & --- & --- & --- & --- \\
        \hline
        & \multicolumn{7}{c}{\bf USco}\\
      0.1--1.0  &       &  --- & --- & --- & 0.24$\pm$0.04 & 0.11$\pm$0.02 & 0.10$\pm$0.01 \\
      1.0--3.0  &     &  --- & --- & 0.37$\pm$0.08 & 0.21$\pm$0.05 & 0.10$\pm$0.01  & 0.10$\pm$0.01  \\
      3.0--9.0& &  --- & 0.59$\pm$0.14 & 0.22$\pm$0.03  & 0.18$\pm$0.01  & 0.20$\pm$0.04 & --- \\
      9.0--30    &     &  --- & --- & --- & 0.19$\pm$0.02 & --- & --- \\
        \hline
        & \multicolumn{7}{c}{\bf Pleiades}\\
        0.1--1.0  &    &  --- & --- & 0.12$\pm$0.01 & 0.08$\pm$0.01 & 0.07$\pm$0.01 & 0.07$\pm$0.01 \\
      1.0--3.0    &   &  --- & --- & --- & 0.06$\pm$0.01 & 0.07$\pm$0.01  & ---  \\
      3.0--9.0  &&  0.04$\pm$0.01 & 0.09$\pm$0.01 & 0.10$\pm$0.01 & 0.05$\pm$0.01  & --- & --- \\
      9.0--30     &    &  --- & --- & --- & --- & --- & --- \\
        \hline
        & \multicolumn{7}{c}{\bf Praesepe}\\
     0.1--1.0   &       &  --- & --- & --- & --- & 0.09$\pm$0.01 & 0.15$\pm$0.02 \\
      1.0--3.0  &     &  --- & --- & --- & --- & 0.08$\pm$0.01  & ---  \\
      3.0--9.0& &  0.03$\pm$0.01 & --- & --- & 0.06$\pm$0.01 & 0.06$\pm$0.01 & --- \\
      9.0--30    &     &  0.04$\pm$0.01 & 0.02$\pm$0.01 & 0.02$\pm$0.01 & 0.04$\pm$0.01 & 0.05$\pm$0.01 & --- \\
        \hline
 \multicolumn{8}{l}{$\star$ Amplitude inferred from disk-bearing stars.}\\
      
         \end{tabular}
      \label{tab:table}
\end{table*}

\section{Conclusions}

We find evidence that the amplitude of photometric variability of low-mass stars decreases with age at constant color and rotation period. At earlier ages, the observed decrease is likely due to the passage from a type of variability dominated by accretion-related phenomena to a variability dominated by magnetic activity phenomena (age range from $<$ 3\,Myr to $\sim$10\,Myr). At older ages, the decrease is likely due to the shrinking of the external convection zone which implies a reduced hydromagnetic dynamo efficiency and a consequent reduced level of magnetic activity. Finally, the lowest mass stars examined in this paper (V$-$K$_s$ $>$ 6 mag) show a hint signaling a reversal in the trend, with a slight increase in the variability  level in the age range from 125\,Myr to 625\,Myr.\\
Independently of the causes, whether related to the time evolution of convective turnover time $\tau_c$ or the above-mentioned ARGD time scale, or both, we can take advantage of this age dependence to constrain the age of low-mass associations and cluster members by using the level of photometric variability. This would be  a statistical approach that requires a number of stars of same  color  and rotation and, thus, is not suitable for individual field stars. 

Indeed, our study has an exploratory value,  especially when considering that we used only five among benchmark associations and clusters. Additional associations and clusters at different ages are needed to indicate which would be the optimal functional relation to describe the age decline of photometric variability.
The coming DR3 release of Gaia data is expected to provide photometric time series for a few tens of thousands of low-mass members of young clusters and associations that span a wide range of ages. The huge amount of data, together with their homogeneity, will be best suited to make use of and further test this newly proposed age diagnostics. \\

{\it Aknowledgments}
Research on stellar activity at the INAF-Catania Astrophysical Observatory is supported by MUR  (Ministero Universit\'a e Ricerca). The Author thanks the Referee for useful comments that allowed to improve the data analysis and overall the quality of the paper. All the data presented in this paper were obtained from the Mikulski Archive for Space Telescopes (MAST). STScI is operated by the Association of Universities for Research in Astronomy, Inc., under NASA contract NAS5-26555. Support for MAST for non-HST data is provided by the NASA Office of Space Science via grant NNX13AC07G and by other grants and contracts. The author thank Dr. Federico Spada for useful discussion in the use of YaPSI database.
SM acknowledges financial support from Progetto Mainstrem “Stellar evolution and asteroseismology in the context of the PLATO space mission” (PI: S. Cassisi).

\bibliographystyle{aa.bst} % style aa.bst
\bibliography{Messina} % your references Yourfile.bib

\end{document}